# Super stealth dicing of transparent solids with nanometric precision

Zhen-Ze Li[1,2,†], Hua Fan[2,†], Lei Wang[1,†,*], Xu Zhang[1], Xin-Jing Zhao[1], Yan-Hao Yu[1], Yi-Shi Xu[1], Yi Wang[2], Xiao-Jie Wang[2], Saulius Juodkazis[3,*], Qi-Dai Chen[1,*] and Hong-Bo Sun[1,2,*]

[1] *State Key Laboratory of Integrated Optoeletronics, College of Electronic Science and Engineering, Jilin University, Changchun 130012, China*

[2] *State Key Laboratory of Precision Measurement Technology and Instruments, Department of Precision Instrument, Tsinghua University, Beijing 100084, China*

[3] *Optical Sciences Centre and ARC Training Centre in Surface Engineering for Advanced Materials (SEAM), School of Science, Swinburne University of Technology, Hawthorn Vic 3122, Australia*

[†] *These authors contribute equally to this work*

[*] *Corresponding authors:*

*Lei Wang: leiwang1987@jlu.edu.cn*

*Saulius Juodkazis: sjuodkazis@swin.edu.au*

*Qi-Dai Chen: chenqd@jlu.edu.cn*

*Hong-Bo Sun: hbsun@tsinghua.edu.cn*

Laser cutting of semiconductor wafers and transparent dielectrics has become a dominant process in manufacturing industries, encompassing a wide range of applications from flat display panels to microelectronic chips. Limited by the diffraction barrier imposed on the beam width and its longitudinal extend of laser focus, a trade-off must be made between cutting accuracy and aspect ratio in conventional laser processing, with accuracy typically approaching a micron and the aspect ratio on the order of $10^2$. Herein, we propose a method to circumvent this limitation. It is based on the laser modification induced by a back-scattering interference crawling mechanism, which creates a positive feedback for homogenizing longitudinal energy deposition and lateral sub-wavelength light confinement during laser-matter interaction. Consequently, cutting width on the scale of tens of nanometers and aspect ratio $10^3 \sim 10^4$ were simultaneously achieved. We refer to this technique as "super stealth dicing", which is validated through numerical simulations, ensuring its broad applicability. It can be applied to various transparent functional solids, such as glass, laser crystal, ferroelectric, and semiconductor, and is elevating the precision of future advanced laser dicing, patterning, and drilling into the nanometric era.

The history of laser cutting, drilling, and patterning dates back to 1965[1] and it has now become an indispensable technology for material removal[2,3], surface texturing[4] and integrated chips[5] in industrial[6,7] and medical applications[8]. Recently, the emergence of time-domain laser modulation, such as ultrafast burst[2], has further enhanced the efficiency and quality of laser processing. Despite decades of development and numerous achievements, achieving nanometric high aspect ratio (AR) cuts remains a formidable challenge for current state-of-the-art laser dicing technologies, for example, width < 100 nm and AR > $10^3$ (see Suppl. Section 1.3, Fig. S4, and Table S1 for a summary of the current state of the art). The underlying reason is that the diffraction-limited wave propagation nature of light contradicts the homogeneous axial and sub-wavelength lateral energy deposition required for nanoscale laser structuring. For any propagating or evanescent electromagnetic waves, the spatial spread of a wavepacket in transverse dimensions $\Delta r_\perp$ links to its angular spread by the bandwidth theorm in Fourier optics[9–11] $\Delta r_\perp \Delta k_\perp \approx 1$, where $\Delta k_\perp$ denotes the bandwidth of transverse spatial frequency. Given the quantized nature of light, this is also equivalent to the famous Heisenberg uncertainty principle in quantum physics[12,13], and has led to the evident fact that a tightly focused beam (small $\Delta r_\perp$) must contain large transverse wavevector components and therefore diverge faster. It enforces a trade-off between the beam width and its longitudinal extend[10,14,15], and raises limitations in laser processing: A mathematically non-diffracting beam requires infinite spread in real space and incident aperture[10,12], making it physically impossible. On the other hand, the large wavevector components carried by optical near fields can provide an ultimate field localization $\Delta r_\perp$ much smaller than the transverse diffraction limit $\approx 1.22\lambda/\text{NA}$ to realize the finest lateral fabrication resolution[16–21]. However, near-field components decay exponentially in the longitudinal direction due to momentum conservation[22], thus cannot provide sufficient processing depth for nanodicing of bulk material (Suppl. Section 1.1). Additionally, the complex light-matter interactions during laser processing[23–25] further exacerbates the cutting/drilling inhomogeneity by linear/nonlinear absorption and scattering[26,27], as well as stress and thermal effects derived from the intense energy deposition[28–32].

To address the challenges above, two mainstream solutions have been proposed. One is the axial-stretched beam by modulating the phase and amplitude of the incident pulse to make a bal-ance between the beam width and its longitudinal extend[33–35]. It can be proved by the Fourier center slice theorem that the angular width $\Delta\theta$ of an annular aperture and the length of diffraction-free region $\Delta z$ of the stretched beam still satisfies a similar conjugate relation: $\Delta z \Delta\theta \approx \lambda/\text{NA}$[10], with $\lambda$ and NA being the laser wavelength and numerical aperture. One of the most well-known and successful examples is the stealth dicing technology[7] based on the Bessel-Gaussian beams[36–38]. Instead of directly cutting the sample, it utilizes laser-induced modification in the bulk of the sample to generate a controlled injection of internal stress, which can trigger the directional propagation of cracks to separate material into pieces under mechanical load. Yet, the only shaping of laser focus is much superficial since a lot of complex light-matter interactions should be considered, drawing off another interesting scheme of laser filamentation[39–41]. Owing to the dynamical balance between the Kerr nonlinearity[5,42] and the defocusing effect by the photoionized plasma, a long-range high-refractive-index channel forms a transient waveguide, thereby avoiding beam divergence caused by diffraction. With these two schemes, the stealth dicing and some associated technologies have demonstrated high-quality micron-size drilling, dicing or patterning on semiconductors/dielectrics such as through-silicon via (TSV)[5] and through-glass via (TGV), which have revolutionized the microelectronic industry[43,44].

Despite these achievements, all these strategies have yielded unsatisfactory results for attempts towards damage-free cuttings, where nanoscale kerf and roughness, zero taper and unlimited aspect ratio are highly anticipated. Intuitively, it is straightforward to decrease the pulse energy to boost the lateral accuracy beyond the diffraction limit, and even down to tens of nanometers simply by the well-known threshold effect[45–49]. However, the threshold effect leads to a reduction in both the lateral dimensions and the longitudinal depth of the processed volume based on the energy distribution of the focal spot, and the final structural aspect ratio is still limited. Thus, although it is possible to achieve microscale dicing of milimeter-thick glass by Bessel beam[50], once the lateral processing resolution reaches to a hundred nanometer scale, the aspect ratio of the repeatable structures rapidly drops around or below $10^2$ and exhibits inhomogeneity in the longitudinal direction (Table S1).

In this article, we demonstrate that the key to solving this problem lies in the self-organization feedbacks of far-field and near-field light-matter interactions in three dimensions, rather than solely manipulating the laser focus profile itself. A conceptually similar but physically different idea of utilizing nonlinear feedbacks can be traced back to the inspiring pioneering works made by B. Öktem[4], O. Tokel[5] and their colleagues, where dipole radiation and self-terminating oxidation were employed to regulate the laser printing of periodic structures on metal surfaces[4], as well as incoherent beam superposition for direct writing of integrated silicon chips[5]. We found that, by utilizing the laser-induced sub-surface nanoseeds as scatters (Fig. 1a), a nanostructure growth mechanism we named back-scattering interference crawling is introduced to constitute the positive feedback for the homogenization of longitudinal energy deposition (see Suppl. Movies). Concretely, the energy redistribution caused by the interference between the back-scattering wave and the incident light field guides the cascade generation of secondary seeds (Fig. 1a, d), while the near-field enhancement drives the elongation, connection and uniformization of these seeds, that is, the "crawling" effect (Fig. 1b, c, and f-h). Through this effect, the local sub-wavelength high-precision structural growth, which is dominated and confined by near-field interactions, can be gradually extended over the entire focal depth of the propagating field (Fig. 1a-d). Furthermore, the robustness of near-field enhancement guarantees the longitudinal uniformity of the final structure (Fig. 2), ignoring the inevitable divergence and inhomogeneity of the input stretched laser beam along the optical axis, as if the constraint from the Fourier bandwidth theorem is "relaxed".

Combined with suitable standard post-processings such as chemical etching[51–53] or stress load[54], nanodicing of various transparent materials are achieved in a width of tens of nanometers, and an aspect ratio $\sim 10^3$ (Fig. 2, Fig. 4b, and Suppl. Section 11.3), as well as the state-of-the-art laser dicing with a pitch of 150 nm and special-shaped drilling/cutting with a sidewall roughness Ra < 10 nm (Fig. 3f-h, Fig. 4, and Fig. 5). Under the optimized processing conditions, the achievable width of the nanocuts can reach $7 \sim 15$ nm with an aspect ratio of $10^4$. As a proof-of-concept, we have utilized this technique to prepare true zero-order sapphire waveplates with retardance accuracy better than ±1 nm, as well as crystalline micro-prisms for beam manipulation. This technology, which we name super stealth dicing (SSD), has significantly improved the resolution and aspect ratio of laser dicing while maintaining the merits of stealth dicing of debris-free and damage-free for in-volume laser processing.

**Mechanism of back-scattering interference crawling** First, we take the fused silica, one of the most important materials in optics and engineering, as an example to discuss the generation of nanoseeds and the subsequent feedback they bring, and the physical pictures therein can be easily generalized to other materials. In the experiments, linear-polarized 230-fs laser pulses at 515-nm wavelength are focused 60 $\mu$m depth below the sample surface with an energy density 10% above the material damaging threshold (details are shown in **Methods** part). Generally, a nanoseed structure arises in the focal center within 5 pulses where the energy is slightly above the damage threshold for the first several pulses. As discussed in Suppl. Section 5, the nanoseed is not only a low-refractive-index void region confirmed by the micro refractive index measurement, but also a dangling-oxygen-enriched area as revealed by Raman spectrum[55–58]. Due to lower photoionization energy compared to its surroundings, the nanoseed will facilitate the local plasma ionization during the subsequent laser irradiation, resulting in a transient permittivity gradient around itself following the Drude-Lorentz model (Fig. 1e and Suppl. Section 2.1). This optical contrast leads to two pronounced feedbacks at far-field and near-field regions. On one hand, the excited nanoseed can be approximated as a dipole, reflecting part of the energy backwards and interfering with the rest incoming laser pulse to form a standing wave (Fig. 1a), which has been verified both theoretically and experimentally (Fig. 1d, f). As a result, there is a periodic modification along the optical axis with periods $\sim \lambda/2n$ ($n = 1.46$ at 515 nm. However, the period is actually affected by the nonlinearity-induced refractive index perturbation, as discussed in detail in Suppl. Section 8.1) and promotes the generation of secondary seeds at new local energy maxima (Fig. 1b, d, and f). Consequently, we refer to this secondary seed generation guided by back-scattering interference, as far-field feedback. On the other hand, the presence of a nanoscale permittivity gradient surrounding the seeds facilitates the excitation of enhanced near fields, allowing them to extend along the optical axis, namely the near-field feedback (Fig. 1b, g, Suppl. Section 3 and 6).

With the further accumulation of laser pulses, the gap among seeds will be reduced by the near-field coupling, being shaped into a connected nanoline (actually it is nanopillar in three dimensions) that finally extends to hundreds of nanometers (Fig. 1b, c, and h). Therefore, the back-scattering interference crawling demonstrates the cooperative nature of far and near fields. It is manifested through the interference that reduces the distance between nanoseeds to sub-wavelength scales, allowing for localized interactions between neighbouring near fields and promoting the connectivity of nanostructures to form a coherent entity. It should be noticed that the presence of longitudinal standing waves necessitates the "coherence" of feedback, ensuring that the feedback range remains within the coherence distance. In contrast, the nonlinear processes for three-dimensional modification in silicon revealed by O. Tokel et al. is a noncoherent superposition of the incident and reflected beams[5].

**Parallel generation of nanoseeds and the assembly of nanoline** The above mechanism provides us with valuable insight, that is, an intrinsic backward crawling effect drives nanoseeds into nanoline if the light intensity is slightly above its threshold. It indicates that as long as we disperse laser focus for seeding over the optical axis simultaneously, a longitudinally-uniform structure will be constructed automatically by the back-scattering interference crawling mechanism, and the divergence and imperfection of laser focus can be bypassed due to the robustness of local-field enhancement. To further clarify this claim, we use the spatial light modulator (SLM) to stretch the beam to tens of micrometres (Fig. 2a, b, and Suppl. Section 7.1 for hologram generation). Just as we expected, seed structures are first generated at those locations where the optical intensity is above the threshold and then induce the secondary seeds in parallel by back-scattering interference (Fig. 2c, d, and e, top lane). In contrast to the condition of Fresnel reflection at interface[37], the back-scattering interference crawling, induced by nanoscale nonlinear permittivity inhomogeneity, can simultaneously occur at arbitrarily designed depths along the optical axis, without the constraint from a specific interface (Suppl. Section 8.5). The contribution from the self-focusing effect can also be excluded considering the low single pulse energy (Suppl. Section 8.7). The situation becomes more complicated when the pulse density is increased to 14 $\mu m^{-1}$. At this stage, all the seed structures are sufficiently elongated and the initial and secondary seeds become almost indistinguishable (Fig. 2e, middle lane). Meanwhile, due to spatial fluctuations in the spatial intensity distribution of the interference field, the spacing of seed structures may become integer multiples of $\lambda/2n$ (Suppl. Section 8.6). When the pulse density reaches 44 $\mu$m$^{-1}$, the structures at the bottom become almost continuous, while the discrete top seed structures are still growing and crawling upward/upstream (Fig. 2e, bottom lane), which also supports our previous interpretation (details on the evolution of seeds can be found in Suppl. Section 8.2, 8.3, and 8.4). In fact, the backward crawling of nanoseeds is also reflected from the linear dependence of seed length on its absolute longitudinal coordinates (Fig. 2g, with a slope of 4.5∼5.5 nm/$\mu$m). For those seeds that are further from the sample surface, they appear earlier and grow longer under pulse accumulation. Finally, when the number of pulses reaches 130 $\mu$m$^{-1}$, the length of the assembled nanoline is ≈ 95 $\mu$m (Fig. 2f).

Here, a critical question arises: What is the maximum allowable depth written by this method? If only one scan is performed, the maximum length depends on the extent that the seed structures can be generated, and is therefore limited by the total length of the light field. However, the back-scattering interference crawling also implies a longitudinal self-alignment mechanism to achieve vertical splicing of nanolines by multiple scans to further breakthrough this limitation. Since a single nanoline can be aligned and combined bottom-up by feedback from discrete nanoseeds (Fig. 2e-h), it is natural to trigger similar feedback to guide the seed generation in subsequent scans from the top of an existing structure (see the simulations and experiments in Suppl. Section 11.2.3). Therefore, the total length of the final structure can theoretically be arbitrarily long without considering the hardware constraint (e.g., working distance of objective lens, range of motion stage). In this regard, the trade-off between lateral accuracy and aspect ratio in laser processing is completely overcome by the back-scattering interference crawling: The resolution and homogeneity of the local nanostructures are guaranteed by the evanescent fields and extended to arbitrary distances by the interference of propagating fields. Conversely, we can also remove a part of the seeds on the optical axis by reducing the radius of the hologram, and then fine-tuning the length of the nanoline (Suppl. Section 11.2.2) without changing the energy of the input laser.

**Extreme sub-wavelength light confinement by nanometric low-permittivity region** As the nanoseeds elongate, their lateral dimension decreases rapidly and converges to a value determined by the laser excitation condition, regardless of its initial size and shape (from Fig. 2e to i). The ultimate width of the nanoline is designable and reproducible and can be as small as $\sim 10$ nm, making the structure aspect ratio approach 10000 in one scan. As shown in Fig. 2**i**, the 10 nm wide laser damage exhibits distinct boundaries in the high-angle annular dark-field (HAADF) mode of the transmission electron microscopy (TEM). The reduced brightness around the nanoline in HADDF, where signal intensity correlates with atomic number contrast, suggests a lower atomic density in the vicinity of the laser-irradiated region. The width of nanolines can also be precisely tuned by chemical etching (Suppl. Section 9). Surprisingly, we found that the formed nanoline also contributes to the uniformization and localization of the light field (Fig. 2i-k). According to the continuity boundary condition for the electric displacement, we have (Suppl. Section 3.1):

$$\frac{\partial E_{\text{pol}}}{\partial x} \propto -E_{\text{pol}} \frac{\partial \ln \epsilon}{\partial x}. \tag{1}$$

The above equation indicates that, for an arbitrary distribution of permittivity along the laser polarization, $\epsilon(x)$, a uniform near field can always be excited and enhanced along the nanoline, as long as the real part of the permittivity around the nanoline is smaller than its surrounding and varies rapidly enough (Fig. 2j, k). In our condition, the transient plasma plays such a role, since it contributes a negative polarization $\Delta \epsilon$ which is proportional to the local plasma density $\rho(x)$. According to classical electrodynamics, the highly localized plasma and light field will further compress transient energy absorption rate since $A(t) \propto \omega \, \text{Im}[\Delta \epsilon]|E(t)|^2 \propto \rho(x)|E(t)|^2$. Correspondingly, all the nonlinear processes occurring in laser fabrication at this stage, ranging from photo-ionization, and inverse bremsstrahlung absorption, to the final optical breakdown, are enhanced and confined within the low-permittivity nanoline. In Suppl. Section 3.2, we numerically simulated this process by self-consistently coupling nonlinear Maxwell's equations and the plasma equations and obtained the spatial distribution of transient absorption rate $A(t)$ can be as low as 25 nm (FWHM, full width at half maxima), which coincides with our experiments well. Figure 2l shows the calculated optical intensity distribution around a nanoline where the field-enhanced re-gion (FWHM $\sim$ 25 nm) becomes consistent with the geometric boundary of the nanoline. Here, the laser governs the creation and evolution of the nanostructures from discrete voids to a continuous line, which in turn modulates and regulates the light field profile to fit the final structure, achieving a sub-20 nm homogenization.

In contrast to the conventional high refractive index nanostructures employed for light manipulation (e.g., waveguide, self-focusing, Mie resonance), this positive feedback loop can be regarded as an interesting analogy of the concept of "light confinement" in nanophotonics, which describes the increase of energy transportation and laser-matter interactions induced by local field enhancement within the low-refractive-index materials[22,59,60]. However, the triggering of this analogue in SSD does not require any artificial structures but emerges naturally from the transient nonlinear interaction that is inherent to the propagation of electromagnetic fields in non-uniform media (Suppl. Section 3.3).

**Self-regulatory formation of nanolines and free-form nanodicing of SSD** The interaction between the nanoseed and the light field in the $xy$-plane is crucial for the followed dynamic laser scanning. We found that the evolution of the nanoseeds in each $xy$-plane can be fully explained by our recently proposed laser direct writing technique named optical far-field-induced near-field breakdown[18], where the rotating far-field polarization guides the dynamical lateral near-field writing. Due to the continuous condition of the electric displacement, the orientation of the near-field enhancement around the nanoline in the horizontal plane is always perpendicular to the laser polarization (Fig. 3a and Suppl. Section 6.1), which results in the anisotropy seed growth in the $xy$-plane (Fig. 3b, c). Accordingly, to achieve optimal control of near-field localization, it is crucial that the translation of the sample should be perpendicular to the linear polarization of the incident beam. Considering a smaller optical contrast ($\delta\epsilon \sim -0.2$ or $\delta n \sim -0.08$, which links to the ambiguities of energy relaxation time when approaching the material breakdown threshold[37,61], see Suppl. Section 2.5) between the excited nanoseed and the host region produced by the laser pulse, the 3D laser writing in the bulk results in a slow expansion of the structure along the polarization direction (Fig. 3c). Moreover, the near-field enhancement is inversely proportional to the nanofeature size, resulting in a self-regulatory nature of the inscribed modification. That is, the nanofeatures will eventually evolve into a uniform size during the scan (within 300 nm, see Fig. 3d, e) regardless of the generation quality or the shape of the initial seeds (Fig. 3e, $0 < d < 100$ nm). This self-regulation effect promotes the high repeatability of the SSD technology. As shown in Figure 3e, the width of nanoslits stablize to $50 \pm 3$ nm when the scan distance to initial seed $d \geq 300$ nm. We further measured the reproducibility of the width of six sequentially written nanolines in Suppl. Section 9.1, which gives an average width of 45.8 nm and a standard deviation of 2.5 nm.

The polarization-controlled near-field fabrication results in a significantly different point of view from conventional laser processing technologies. Traditionally, two parameters of scanning speed and laser repetition rate are accustomed to being combined to estimate the energy deposition per unit length and establish a connection with the experimental results. However, here the far-field beam is not really taking part in processing until it is converted into near fields by the nanoseeds. Accordingly, the spacing between adjacent pulses becomes the core parameter in SSD technology, as it modulates the overlap ratio of far-field beam and nanoseeds, which totally determines the morphology of near-field hot spot (Suppl. Section 8.3). In our experiments, a step size of 20 nm and $2 \sim 5$ laser pulses per step were taken to fully exploit the self-regulatory evolution of the near fields. Owing to the highly localized near-field energy absorption, the thermal accumulation can be effectively avoided in SSD technology, and the processing results do not change observably with the repetition frequency of the laser pulses within the range of 200 kHz (Suppl. Section 12). Accordingly, the pulse density mentioned above leads to a scanning speed of 0.1 mm/s for 10 kHz and 2 mm/s for 200 kHz.

To demonstrate the superiority of SSD technology, we show that high-AR nanodicing with any designed period and duty cycle can be achieved (see **Methods** and Suppl. Section 9 for detailed sample preparations). As shown in Figure 3f-h, we can observe part of the cross-section of these gratings through ion beam milling. Limited by the milling depth of the ion beam, we cannot show the structure along its entire depth. The grating with period of 300 nm and a duty ratio of 50%, its width is 150 nm and maintains good longitudinal uniformity within the visible 30 $\mu$m depth. To the best of our knowledge, this represents the state-of-the-art quality of nanoscale high aspect ratio structures obtained by laser direct writing.

**General applicability of SSD** Now we can extend the established results in fused silica to other transparent materials. The success of this promotion lies in the fact that triggering the backscattering interference crawling only requires the nanoseed itself to act as a low permittivity region (compared to its surroundings), which can be easily satisfied by the laser-damaged region. For instance, extensive studies on fused silica have indicated that plasma excitation during laser irradiation, as well as the defect generation and local density reduction after irradiation, all lead to a decrease in the permittivity (Suppl. Section 5 and 8.1). Similarly, in crystals, the crystalline-to-amorphous phase transition after laser irradiation can also result in a lower permittivity. As an evidence, we also show the existence of nanoscale low-refractive-index regions in YAG crystals after laser irradiation through micro refractive index measurement and high-resolution transmission electron microscope (Suppl. Section 5).

As a result, we can successfully extend nanodicing to various functional crystals with different bandgaps, refractive indices, and symmetries (Fig. 4a-c, the processing parameters can be found in Suppl. Section 11.1.2, Table. S5 and S7). These materials include lithium tantalate[62], lithium niobate (ferroelectric and nonlinear optical applications[63]), YAG and its Ce-doped variants (lasing, radiation detection, and high-power LED[64]), Ti: sapphire, and n-type doping $\beta$-$Ga_2O_3$ (a fourth-generation semiconductor with applications in high-power devices and ultraviolet detectors[65,66]). Similarly to fused silica, we have also observed the transition from discrete nanoseeds to continuous nanolines in YAG and lithium niobate, which provides additional confirmation of the generality of our theory in crystalline materials (Suppl. Section 11.2.2 and 11.3.3). Notably, the narrowest dicing width obtained in lithium niobate crystal can reach 7 nm without the need of chemical etching, which equals only 14 lattice constants ($a = 5.148$ Å, Suppl. Section 11.3.1). To dispel the possibility of the 7 nm cuts in lithium niobate being superficial defects of only a few nanometers deep, we conducted additional sample characterization using side polishing and wet etching. As elaborated in Suppl. Section 11.3.2, nanodicing in lithium niobate can reach depths of up to 230 $\mu$m, with a characteristic width of 15 nm, resulting in an aspect ratio exceeding $10^4$. In should be noticed that, apart from cubic YAG crystal, the other crystalline materials fabricated in this study, are uniaxial or biaxial crystals and exhibit natural birefringence. This causes the polarization state to continuously evolve on the Poincaré sphere[67], deviating from strict linear polarization at the laser focus. Such behaviour may potentially compromise the effectiveness of near-field enhancement. Fortunately, through a combination of theoretical calculations and experiments, we have demonstrated the robustness of near-field enhancement in SSD against weak birefringence perturbations, ensuring its functionality (detailed analysis and experiments on $r$-plane sapphire, $\beta$-$Ga_2O_3$ and $YAlO_3$, refer to the Suppl. Section 11.1).

As shown in Fig. 4d-h, we have further prepared nanorods (800 nm period) in Ce: YAG crystal and nanowires in Ti: sapphire crystals. These nanoscale structures exhibited highly smooth surfaces under optical, electronic and atomic force microscopes, and can be manipulated freely by a nanomanipulator. Furthermore, we demonstrated fluorescence mapping of a 0.5 $\mu$m × 20 $\mu$m Ti: sapphire nanowire oriented along the $c$-axis, and its corresponding fluorescence spectrum (transferred to PDMS substrate, Fig. 4i). The fluorescence intensity of the nanowire showed a significant dependence on the polarization orientation of the excitation laser (488 nm, see **Methods** for details). When the excitation laser was polarized along the long axis of the nanowire ($\pi$-polarization, parallel to the $c$-axis), the fluorescence intensity was approximately 1.5 times higher than when the excitation light was perpendicular to the nanowire ($\sigma$-polarization). This observation is consistent with the property of bulk Ti: sapphire, further confirming the non-destructive property of the SSD technique. The achieved crystal cutting results hold promising applications in fields such as nanoscale laser[22,68] and sensing[69] in the future.

Through SSD technology, we are able to perform direct cuts on birefringent crystals along specific optical axes, enabling the fabrication of true zero-order waveplates with customized wavelengths. Fig. 4j shows the ultra-thin sapphire slices cut from the A-plane sapphire ($\langle 1\bar{1}20 \rangle$) along the $c$-axis ($\langle 0001 \rangle$, marked by the pseudo color). When the thickness of the sapphire is varied continuously from 4 to 11 $\mu$m, the retardance of the slices vary continuously from 35 to 90 nm with standard deviation < 1 nm. A linear regression yielded a birefringence value of $\Delta n = |n_o - n_e| = 0.0082 \pm 0.0002$ at 633 nm. Based on these results, we further prepared a sapphire 1/4 waveplate with a thickness of 19.855 $\mu$m (measured by confocal microscope) which works at a central wavelength of 660 nm to convert linearly polarized light to circular polarization. The performance of 1/4 wavepalte is tested by a broad, linearly polarized, white-light continuum (460 to 800 nm, see **Methods**), and the ellipticity of the polarization state achieves its minimum value of 45.33° at 660.7 nm, which is only 0.7 nm deviation from the design wavelength (Fig. 4**k**). Similarly, we prepared 1/4 waveplates with center wavelengths of 610 nm and 560 nm, with wavelength deviations of only 0.5 nm and 1.2 nm (Fig. 4**i**).

**Precise specially-shaped drilling of ultra-thin glass by SSD** The SSD technology is not only suitable for high-precision nanoscale dicing, but also can be easily extended to mesoscale precise drilling, which is crucial for bio-medical and microelectronics. Despite that the SSD technique can also realize high-quality cutting of ultra-thin glass and sapphire by applying stress only (analogue to traditional stealth dicing, Suppl. Section 10.2), the utilization of chemical etching offers more flexible patterning capabilities. Here we demonstrate specially-shaped holes in silica glass with a thickness of 40 $\sim$ 70 $\mu$m by the following two steps. First, sub-100 nm modifications/cracks are fabricated on the ultra-thin glass by the method discussed above. Then, wet etching was adopted to widen the gap to $\sim$ 400 nm (Suppl. Section 10.1) since the hydrofluoric acid solution can react with fused silica even without laser modification, so that the block surrounded by the cutting edge falls off. Since the etched area is precisely defined by the modification, geometric distortion including rounded corners caused by isotropic etching is minimized (Suppl. Section 10.1, Fig. S45). As shown in Figure 5a, b, and c, scalable precise drillings with sharp edges can be obtained for arbitrary shapes. Microscale silica cuboids can be collected by filter paper after etching (Suppl. Section 10.1, Fig. S46), which are smooth when observed by an optical microscope (Fig. 5d and inset). To support this observation, we selected an area on the cutting surface of the silica cuboids with a dimension of 40 $\times$ 6 $\mu$m$^2$ for AFM scanning (the optical image of scanning area can be found in Suppl. Section 10.3). As shown in Fig. 5**e**, the one-dimensional height profile shows that most of the height fluctuations lie in the interval of $\pm$30 nm, which cooresponds to a roughness (Ra, defined by the mean absolute deviation) of 9.4 nm. Therefore, the surface roughness achieved through SSD technology is only a fraction of the visible light wavelength ($\lambda/50$ at $\lambda = 532$ nm), which further confirms our observations under the optical microscope.

Due to the high dicing quality of SSD technology, we can cut a variety of optical elements directly from bulk crystals. Traditionally, the typical size of current crystalline prisms is still confined to the millimeter scale due to limitations of conventional machining methods. As a proof-of-concept, we have fabricated dispersive prisms, pentagonal prisms, and deformation prisms with dimensions in the tens of micrometers range using YAG crystals (Fig. 5f). The geometrical profile of these prisms is precisely defined by the continuous laser scanning path, resulting in minimal processing errors (Fig. 5g). Additionally, we prepared right-angle prisms and Fresnel rhombs, conducting subsequent tests to evaluate their performance (see Methods for experimental set-up). As depicted in Fig. 5h, when we focus a 532-nm Gaussian beam to a diameter of 50 $\mu$m and direct it from the right angle of the prism, the prism effectively utilizes total internal reflection at the hypotenuse to deflect the beam by 90°, while preserving a Gaussian spatial distribution in the outgoing beam. Similarly, in the case of the Fresnel rhomb shown in Fig. 5i, a 532-nm laser beam polarized at a 45-degree angle and incident perpendicularly onto the surface will undergo two total internal reflections at a designed angle of incidence of 34.583°, creating a phase difference of $\pi/2$ between $s$-polarized and $p$-polarized components, thereby resulting in circularly polarized light as the output. As shown in Fig. 5j, the ellipticity of the output circular polarization is 44.64°(orange solid), which is very close to the idea condition (ellipticity = 45°, gray dashed). These small prisms can be further utilized in the future for integrated atomic chips[70], bio-imaging[71] and on-chip free-space lasers[72], and to further reduce the dimensions of the devices.

**Conclusion** The super stealth dicing (SSD) for robust free-form nanodicing of transparent solids is presented here with cut widths smaller than tens of nanometers and aspect ratios of $10^3 \sim 10^4$. The principle of such a sharp and accurate 3D nanoknife is empowered by the back-scattering interference crawling and the near-field-induced light localization controlled by far-field polarization, which ensures the applicability of SSD technology for various transparent materials and desired dicing depths if the laser energy is sufficient to support a required extent of the focus modulation[50], or by simply employing multiple scans. This method may open a new path to integrated optics, optoelectronics, fibre-optics, and microelectronics[68,72–74].

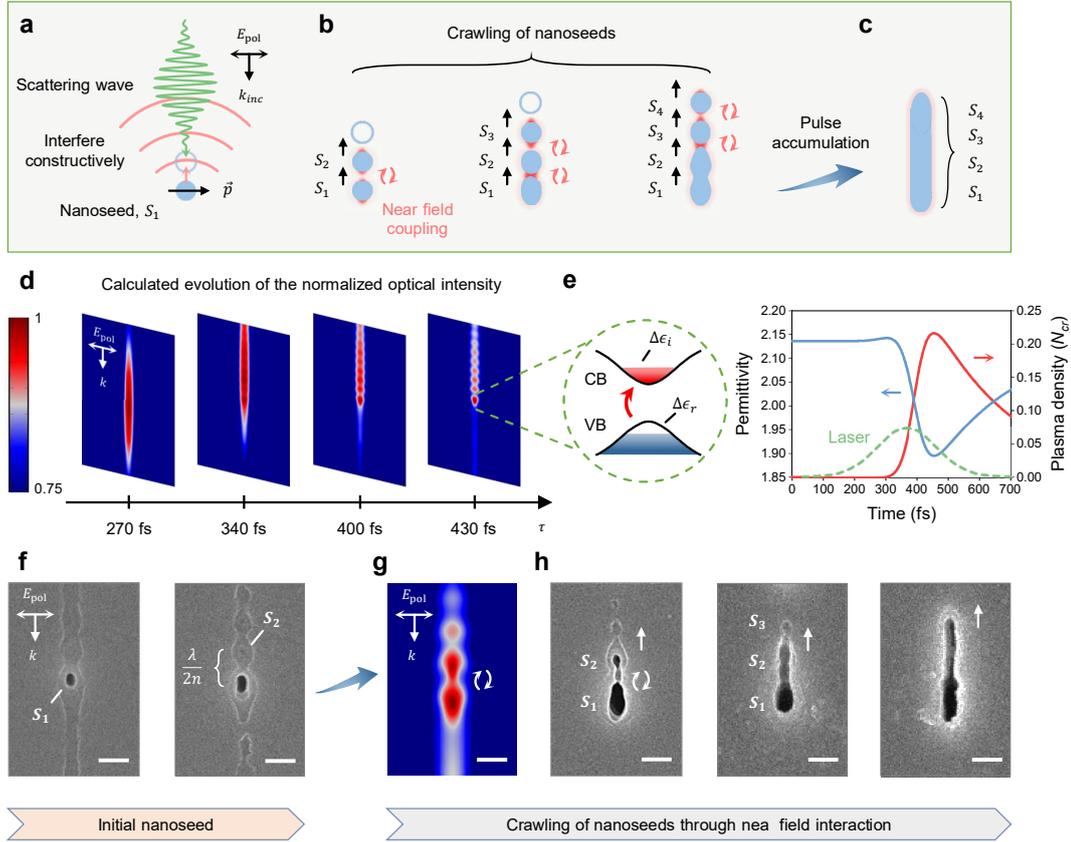

**Fig. 1|Mechanism of the back-scattering interference crawling.** **a-c** Schematic of the nanoseeds backward crawling during pulse accumulation. $E_{\text{pol}}$ represents the linear polarized electric field and $k_{inc}$ denotes the propagation direction of incident laser beam. **d** Calculated time evolution of the normalized optical intensity when a single nanoseed is excited to trigger the back-scattering interference. All the color scales in the main text are set between 0.75 and 1 for better visualization. **e** Schematic and simulations of the transient permittivity modification (for unexcited silica, $\epsilon = n^2 = 2.13$) by plasma excitation. The time envelope of the laser pulse is indicated by the green dashed (not to scale), which has a form of $E(t) = E_0 \exp -[(t-t_0)/\sigma]^2$ with pulse centred at $t_0 = 342$ fs, $\sigma = \mathrm{d}\tau/2\sqrt{\ln 2} = 98$ fs and $\mathrm{d}\tau = 230$ fs is the pulse width as FWHM. **f** Image of the initial seed $S_1$ and the secondary seed $S_2$ induced by the back-scattering interference. **g** Calculated normalized optical intensity around two adjacent nanoseeds along the optical axis. **h** Experimental verifications of the near-field coupling and the nanoseeds crawling under the subsequent pulse irradiations. All the in-volume nanoseeds are mechanically polished to the surface and etched in 2% hydrofluoric acid solution for 25 seconds for better visibility of the nanostructures in scanning electron microscopes (SEM). All the scale bars are 200 nm.

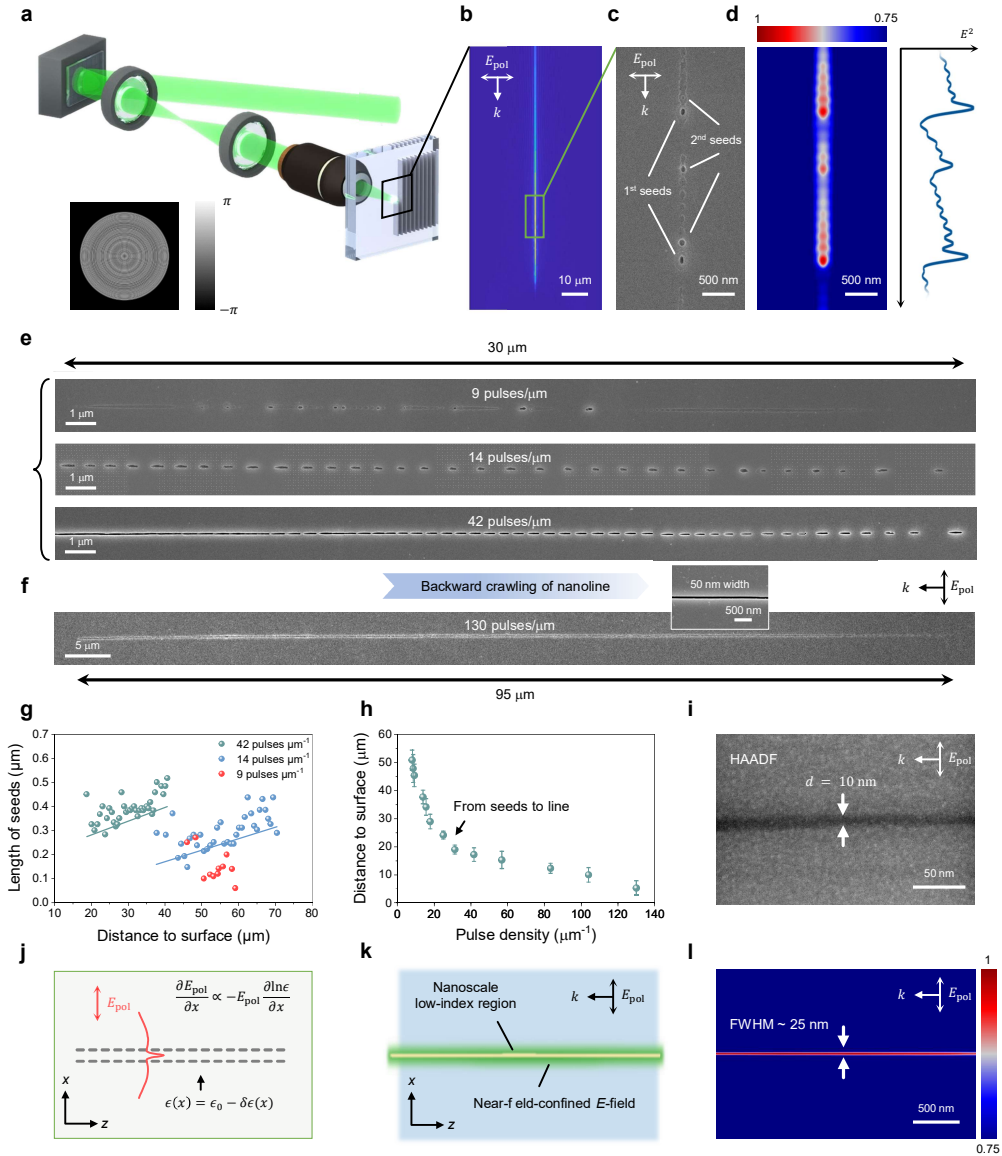

**Fig. 2 | Transition from separated nanoseeds to a nanoline.** **a** Experimental setup for the generation of a stretched beam. **b** Calculated far-field beam profile of the stretched beam. **c** Parallel generation of the initial nanoseeds and the subsequent cascade generation of secondary seeds along the optical axis. **d** Calculated modification of the optical intensity by the existing nanoseeds. **e-f** Evolution from nanoseeds to nanoline with increased pulse density. The growth direction of the nanostructure is opposite to that of the incident laser. The inset SEM image shows that the typical width of nanoline is less than 50 nm. **g** Statistics of the longitudinal length of nanoseeds. The solid line indicates a slope of 5 nm/$\mu$m. **h** Dependence of distance between the top of the nanostructures and sample surface to the pulse density. The black arrow indicates the transition from separated nanoseeds to nanoline. **i** TEM image (HAADF) of a single nanoline formed in fused silica. The nanoline is not etched, which shows the ultimate line width can be $\sim$ 10 nm. **j-k** Schematic of the light confinement via nanoscale low permittivity region. **l** Simulation of the optical intensity distribution confined by a nanoline, the full-width-at-half-maximum of the near field is $\sim$ 25 nm.

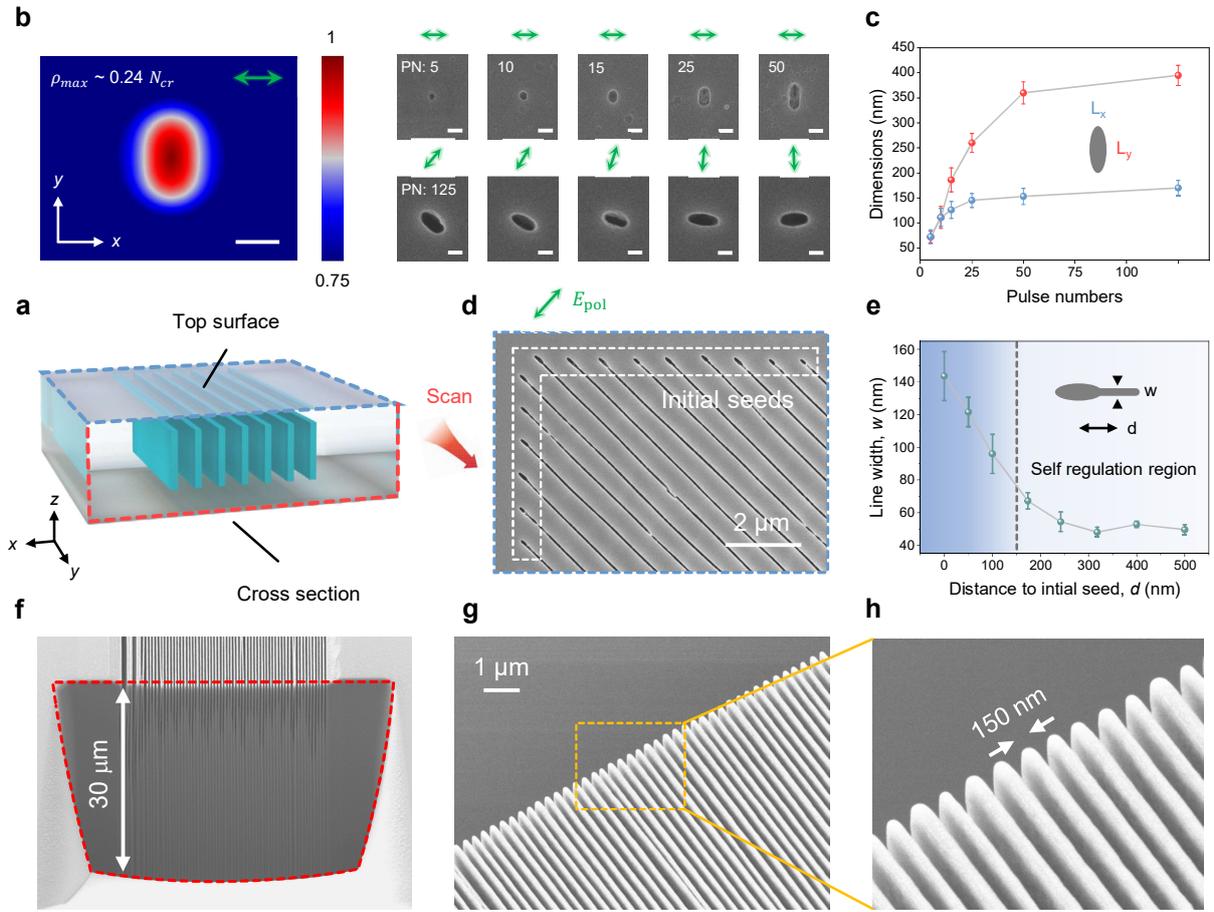

**Fig. 3 | Self-regulatory formation of nanolines and free-form nanodicing by SSD.** **a** Schematic of the 3D characterization of nanocutting. **b** Simulation and experiments of the anisotropy growth of nanoseeds (perpendicular to the laser polarization, indicated by green arrows) in $xy$-plane due to the near-field enhancement. PN represents the accumulated pulse number for each nanoseed. The color scale is set between 0.75 and 1 for better visualization. All the scale bars are 200 nm. **c** The growth of nanoseeds with the increased PN. **d-e** Self-regulatory effect in $xy$-plane. **f-h** SEM images and close-up-view the of nanogratings of a period $\sim$ 300 nm. The collapse of the nanograting in **f** is due to capillary forces after wet etching, and the blurring at the bottom of the structures is due to the redeposition effect during the ion beam milling. The nanogratings were etched by 2% hydrofluoric acid solution for 300 seconds. The pulse density used in **f-h** is 250 $\mu m^{-1}$.

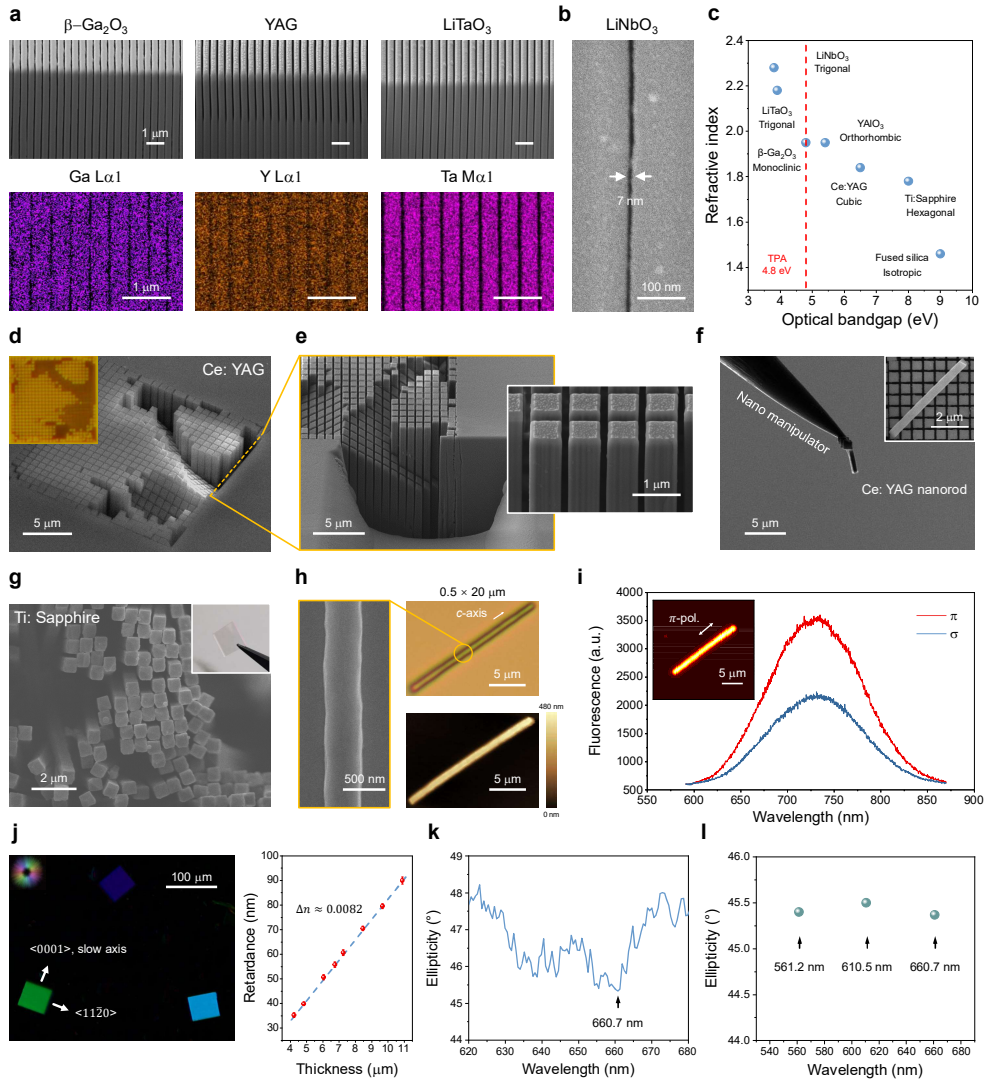

**Fig. 4 | Applicability of SSD technology for various functional materials.** **a** Cross sections of nanodicing revealed by ion beam milling and its corresponding EDS element imaging. The period of nanodicing is 400 nm. All scale bars are 1 μm. **b** 7 nm nanoslit fabricated in $LiNbO_3$ crystal before chemical etching. **c** Physical properties (bandgap vs refractive index) of the fabricated materials by SSD method. The red dashed indicates the energy required for two-photon absorption (TPA) at 515 nm. **d-f** Nanodicing of 800-nm rods in doped crystal Ce: YAG which can be further manipulated freely by a nanomanipulator. Inset: A nanorod on the diced 600 nm period grid. **g-h** Images of the optical, electron, and atomic force microscopes of Ti: sapphire nanowires (c-axis orient, 0.5×20 μm). **i** Mapping of the fluorescence intensity of a Ti: sapphire nanowire (excited by 488 nm continuous laser), which exhibits a strong polarization dependence π-polarization for parallel to c-axis and σ for perpendicular to c-axis of the excitation laser). **j** Birefringence imaging of sapphire ultrathin waveplates, and modulation of the retardance with an accuracy of ±1 nm achieved by precise control of the thickness of the sapphire slice. The working wavelength of the birefringence microscope is 633 nm. **k** Changes of the ellipticity with wavelength of the circularly polarized light modulated by a 1/4 true zero-order sapphire waveplate at 660.7 nm. **l** Fabrication of 1/4 waveplates at designed wavelength with wavelength error less than 1.2 nm.

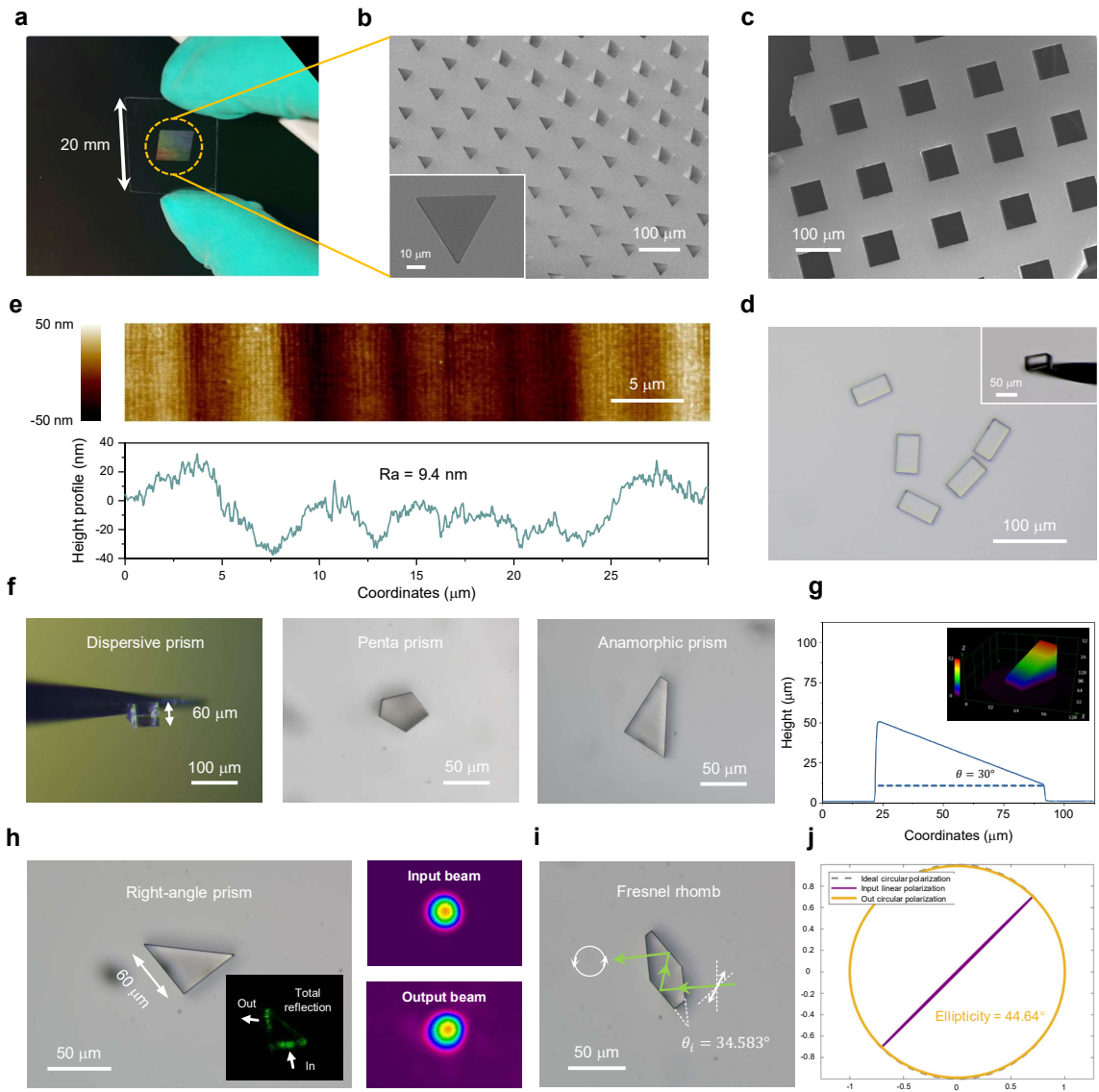

**Fig. 5 | Precise specially-shaped drillings by SSD**. **a-c** Specially-shaped drilling and its close-up-view. The thickness of the ultra-thin silica glass is 70 $\mu$m. The samples were etched by 5% hydrofluoric acid solution for 300 seconds. **d** Optical microscope of microscale silica rods collected after wet etching. **e** AFM characterization of the cutting edge, which exhibits a roughness Ra = 9.4 nm and a relief height of the incision within ± 30 nm over the 30 $\mu$m measurement range. **f** Optical microscope images of the micro-prisms directly cut from the YAG crystal. **g** The height profile of the anamorphic prism obtained through confocal microscopy. The angle between the hypotenuse and the base of the prism is 30°. **h** Optical microscope image of right-angle prism and the measured profiles of the incident and reflected beams. **i-j** The Fresnel rhomb and its performance characterization, which convert normally incident 45-degree linearly polarized light (ellipticity = 0.42°, purple line) to circularly polarized light (ellipticity = 44.64°, orange line). The laser wavelength used in **h-j** is 532 nm and the details for experimental set-up can be found in **Methods**.

**Methods**

**Laser fabrication system** Femtosecond (fs) laser was used for fabrication of samples (Pharos, Light Conversion). The laser delivers $t_p = 230$ fs pulses at $\lambda = 515$ nm with a maximum repetition rate of 200 kHz. The beam diameter was expanded by a beam expander with a magnification of 3.2x. A spatial light modulator (LCOS-SLM, X15213-16, Hamamatsu) was used to impose onto the fs-laser beam a lateral phase pattern for beam stretching in this study. Laser beam modified by SLM were projected by a $4f$ system (750 mm and 300 mm for focal lengths, reducing the beam diameter by 2.5x) and were tightly focused inside the dielectric with a high numerical aperture objective lens (MPLFN, NA = 0.9, Olympus). A three-dimensional piezo stage (P-561.3CDP, Physik Instrumente) that can provide 100 $\mu$m close-loop travel range for each axis was used for nano-positioning during laser fabrication. A three-dimensional step motor stage (MTMS102 for $xy$-plane motion and MVS303 for $z$-axis motion, Beijing Optical Century Instrument) was used for coarse movement and alignment. The energy for laser processing is modulated by a combination of a zero-order $\lambda/2$ waveplate and a polarizing beamsplitter (PBS). The laser polarization was rotated by another zero order $\lambda/2$ waveplate (LUBANG Technology). The wave plate is driven by a stepper motor rotary stage (MRS312, Beijing Optical Century Instrument).

**Beam profile** The shape of the input beam has been measured by beam profilers (Newport, LBP2 series and Coherent, LaserCam HR II). The incident beam width for laser processing is equal to 3.863 mm, which is defined as four times the standard deviation of its intensity distribution ($D4\sigma$). The detailed discussion are shown in Suppl. Section 7.3.

**Sample preparation** Fused silica was procured from Corning (7980STD). Pure single crystals of YAG ($\langle 100 \rangle$), LiNbO$_3$ ($\langle 0001 \rangle$), LiTaO$_3$ ($\langle 0001 \rangle$), sapphire ($\langle 1\bar{1}02 \rangle$), and YAlO$_3$ ($\langle 010 \rangle$) of $10 \times 10 \times 0.5$ mm$^3$ were procured from HF-Kejing Material Technology. Doped crystals of $\beta$-Ga$_2$O$_3$ ($\langle 100 \rangle$, $5 \times 5 \times 0.65$ mm$^3$, 0.03 mol% n-type doping concentration of Sn atom with a resistivity of $0.05 \sim 0.15$ $\Omega \cdot$cm), Ce:YAG ($\langle 100 \rangle$, $10 \times 10 \times 0.5$ mm$^3$, 0.2% doping) and Ti:sapphire ($\langle 0001 \rangle$, $10 \times 10 \times 1$ mm$^3$, 0.1% doping) were procured from CASCRYSTECH Co., Ltd.

The processing strategy is universal for all samples. The stage travels perpendicularly to the laser polarization and a scanning step of 20 nm is set for the piezo stage to fully empower the self-regulatory evolution of near-field fabrication. The experimental results in this article are independent of the laser repetition rate ($\leq$ 200 kHz in the main text). Therefore, we adopt the absolute number of pulses within a unit length, known as pulse density, to characterize the evolution of structures during laser processing. A pulse density of 250 $\mu m^{-1}$, which represents 5 shots for each 20 nm step was used for nanodicing. The pulse density can be transformed to the scanning speed at a specific laser repetition according to the formula below:

$$s(\text{mm/s}) = \frac{\text{laser repetition (Hz)} \times \text{step size (mm)}}{\text{shots per step}}. \qquad (2)$$

A single pulse energy of 385 nJ was used for nanoline inscription inside fused silica (focal depth $d = 60$ $\mu$m) and a single pulse energy of 420 nJ was adopted for the specially-shaped drilling of ultra-thin silica glass. The detailed material's properties and fabrication parameters can be found in Table S6 and S7 (Suppl. Section 11.1.2).

**Threshold measurement** In typical laser surface processing, the Liu's method is commonly used to determine the damage threshold (fluence, $J/cm^2$). However, the laser-induced damage in this work is extended in three dimensions. Accordingly, the threshold is replaced by the minimum energy density ($J/cm^3$, total energy divided by the estimated focal volume) required for observable damage under the optical microscope. The measured threshold for in-volume damage of fused silica $\approx 9025$ $J/cm^3$ (detailed measurement and calculation can be found at Suppl. Section 4). Accordingly, the energy density for nanodicing of fused silica is $10058$ $J/cm^3$ (385 nJ), which is 11% above the threshold.

**Hologram design and focal depth** The function of hologram is to produce an elongated laser focus by using the spherical aberration at the air-sample interface. When the laser beam is stretched, nanoseeds can be spread along the optical axis and trigger the back-scattering interference crawling simultaneously at different depths along the optical axis. Details on the hologram design and focal depth selection are shown in Suppl. Section 7.1 and 7.2.

**Sample characterization** For characterization of in-volume nanostructures, mechanical polishing (UNIPOL-802, SY-Kejing Auto-instrument) and FIB milling (ZEISS Crossbeam 540) were used to expose the profile of three-dimensional nanostructures. For an easier recognition of laser-structured regions in SEM, wet etching of fused silica in 2% HF aqueous solution was implemented considering that the defects in the laser-modified region exhibit a stronger chemical activity[51,75]. Detailed information about etching parameters for other materials can be found in Table S5 (Suppl. Section 9). A field-emission scanning electron microscopy (SEM, JSM-6700F, JOEL) was used for structural imaging of the laser-modified regions. An atomic force microscope (AFM, Dimension Icon, Bruker) was used for the roughness characterization of laser dicing. High-resolution transmission electron microscopy (HR-TEM) of laser-damaged YAG crystal was performed by JSM-2100 (JOEL). The HAADF image of laser-damaged fused silica was performed by FEI Talos 200X. The photoluminescence of a Ti: sapphire nanowire ($0.5 \times 20$ $\mu$m, $c$-axis orient) was tested by a fluorescence confocal microscopy. A linear-polarized 488 nm continuous laser was focused ($100\times$, NA $= 0.95$) to excite the nanowire, and the fluorescence from the PDMS substrate was excluded by a 550 nm long pass filter.

**Optical set-up for performance tests of the waveplates and micro-prisms** The white-light continuum (460 to 800 nm) is generated by a 5 mm thick sapphire crystal pumped by a focused femtosecond laser beam (100 fs, $\lambda = 800$ nm). The white light is focused by the objective lens ($20\times$, NA$= 0.3$) to the sapphire waveplate, and the transmission spectrum is collected by a lens ($f = 100$ mm) and measured by a spectrometer. A rotatable linear polarizer is placed in front of the spectrometer to detect the intensity of polarized components along different directions. A lens with a focal length $f = 50$ mm is used to focus a 532-nm Gaussian beam onto the micro-prism normally. Given the beam diameter $r = 1$ mm, the waist $w$ of the focused Gaussian $d$ beam can be estimated by the following formula $w = \lambda f/r = 25.76$ $\mu$m, which corresponds to a Rayleigh length of 4.05 mm. Within the Rayleigh length, the divergence of the Gaussian beam is negligible and it will undergo total internal reflection at the interface of the right-angle prism. The reflected beam was focused by a lens ($f = 125$ mm) and captured by the beam profiler. From Fresnel's equations, it can be known that the relationship between the angle of incidence $\theta_i$, crystal refractive index $n_{\text{crystal}}$ and phase shift $\delta$ satisfies the following formula:

$$\delta = 2\arctan\left[\frac{\cos\theta_i\sqrt{\sin^2\theta_i - n_{\text{crystal}}^{-2}}}{\sin^2\theta_i}\right], \tag{3}$$

where $\delta = \pi/2$ and $n_{\text{crystal}} = 1.838$ ($\lambda = 532$ nm) imply that an angle of Fresnel rhomb $\theta_i = 34.583°$ is required to generated a circularly polarized beam. The circular polarization was measured by a polarimeter (PAX1000VIS,Thorlabs).

**Statistic and error bars** Due to the random nature of the location and geometric size of the nanoseeds, we need to take a statistical approach to give a description of the range of measured quantities. In Fig. 2h, Fig. 3c and Fig. 3e, error bars are defined by the standard deviation $S = \sqrt{\frac{\sum_i^n (x_i - \bar{x})^2}{n-1}}$ of the quantity $x$ measured by repeating the experiment $n$ times, where $\bar{x} = \frac{\sum_i^n x_i}{n}$ is the average value. In practice, we set $n = 8$ to perform the calculation. For each thickness of the waveplates shown in Fig. 4j, we take 8 samples and 50 points for each sample to measure the retardances by a birefringence microscope (light source $\lambda = 633$ nm) to calculate the standard deviation.

**Numerical simulations** COMSOL Multiphysics 5.6 was utilized for the numerical simulations. Nonlinear Maxwell equations[76] were coupled to the dynamical rate equations[77] to calculate the nonlinear optical response of nanoseeds and its feedback to the light field. The related discussions on numerical simulation can be found in the Suppl. Section 2.4.

**Acknowledgements** This work was supported in part by the National Natural Science Foundation of China (NSFC) under Grants #61825502, #61960206003, #61827826, #62175086; Key Research and Development Program of Shandong Province 2021CXGC010201; Natural Science Foundation of Jilin Province 20220101107JC. S.J. is grateful for the Australian Research Council DP240103231 grant. Z.-Z.L. would like to express gratitude to Prof. Xian-Bin Li, Dr. Yu-Ting Huang, Dr. Jing-Chen Zhang, and Dr. Feng Yu for their valuable discussions. Z.-Z.L. thanks to Dr. Zi-Wen Ma and Dr. Hong-Lin Zhang for their assistance with AFM measurements. Additionally, Z.-Z.L. acknowledges Dr. Yuhao Lei and Dr. Le-Yi Zhao for their support in conducting tests on sapphire waveplates.

**Author contributions**   Z.-Z.L., H.F., L.W., S.J., and H.-B.S. conceived the experiments. Z.-Z.L., H.F., X.-J.Z, and X. Z. carried out the experiments. Z.-Z.L. and H.F. performed the numerical simulations. Z.-Z.L., L.W., H.F., Q.-D.C., S.J. and H.-B.S analyzed the data and improved the results. H.F., Y.-H.Y., Y.-S.X, Y.W., and Z.-Z.L. developed and improved the fabrication system. L.W., Q.-D.C., S.J., and H.-B.S. supervised the whole project. Z.-Z.L., L.W., and H.-B.S. wrote the initial draft and all authors contributed to the final paper.

**Competing financial interests**   The authors declare that they have no conflict of interest.

**Additional information**   Supplementary Information is available for this paper. Correspondence and requests for materials should be addressed to L. Wang, S. Juodkazis, Q.-D. Chen or H.-B. Sun.

**Reference**

1. Adams, C. M. & Hardway, G. A. Fundamentals of laser beam machining and drilling. *IEEE Transactions on Industry and General Applications* 90–96 (1965).

2. Kerse, C. *et al.* Ablation-cooled material removal with ultrafast bursts of pulses. *Nature* **537**, 84–88 (2016).

3. Park, M., Gu, Y., Mao, X., Grigoropoulos, C. P. & Zorba, V. Mechanisms of ultrafast GHz burst fs laser ablation. *Science Advances* **9**, eadf6397 (2023).

4. Öktem, B. *et al.* Nonlinear laser lithography for indefinitely large-area nanostructuring with femtosecond pulses. *Nature photonics* **7**, 897–901 (2013).

5. Tokel, O. *et al.* In-chip microstructures and photonic devices fabricated by nonlinear laser lithography deep inside silicon. *Nature Photonics* **11**, 639–645 (2017).

6. Malinauskas, M. *et al.* Ultrafast laser processing of materials: from science to industry. *Light: Sci. Appl.* **5**, e16133 (2016).

7. Kumagai, M. *et al.* Advanced dicing technology for semiconductor wafer—stealth dicing. *IEEE Transactions on Semiconductor Manufacturing* **20**, 259–265 (2007).

8. Wang, C., Fomovsky, M., Miao, G., Zyablitskaya, M. & Vukelic, S. Femtosecond laser crosslinking of the cornea for non-invasive vision correction. *Nat. Photon.* **12**, 416 (2018).

9. Novotny, L. & Hecht, B. *Principles of nano-optics* (Cambridge university press, 2012).

10. Indebetouw, G. Nondiffracting optical fields: some remarks on their analysis and synthesis. *JOSA A* **6**, 150–152 (1989).

11. Chávez-Cerda, S. A new approach to Bessel beams. *Journal of modern optics* **46**, 923–930 (1999).

12. Durnin, J., Miceli Jr, J. & Eberly, J. H. Diffraction-free beams. *Physical Review Letters* **58**, 1499 (1987).

13. Bialynicki-Birula, I. & Bialynicka-Birula, Z. Heisenberg uncertainty relations for photons. *Physical Review A* **86**, 022118 (2012).

14. McCutchen, C. Generalized aperture and the three-dimensional diffraction image. *JOSA* **54**, 240–244 (1964).

15. Wong, L. J. & Kaminer, I. Abruptly focusing and defocusing needles of light and closed-form electromagnetic wavepackets. *ACS Photonics* **4**, 1131–1137 (2017).

16. Betzig, E. & Trautman, J. K. Near-field optics: microscopy, spectroscopy, and surface modification beyond the diffraction limit. *Science* **257**, 189–195 (1992).

17. Liao, X. *et al.* Desktop nanofabrication with massively multiplexed beam pen lithography. *Nature communications* **4**, 2103 (2013).

18. Li, Z.-Z. *et al.* O-FIB: far-field-induced near-field breakdown for direct nanowriting in an atmospheric environment. *Light: Science & Applications* **9**, 1–7 (2020).

19. Plech, A., Leiderer, P. & Boneberg, J. Femtosecond laser near field ablation. *Laser & Photonics Reviews* **3**, 435–451 (2009).

20. Lei, Y. *et al.* High speed ultrafast laser anisotropic nanostructuring by energy deposition control via near-field enhancement. *Optica* **8**, 1365–1371 (2021).

21. Chen, L., Zhou, Y., Li, Y. & Hong, M. Microsphere enhanced optical imaging and patterning: from physics to applications. *Applied Physics Reviews* **6**, 021304 (2019).

22. Wu, H. *et al.* Photonic nanolaser with extreme optical field confinement. *Physical Review Letters* **129**, 013902 (2022).

23. Jiang, L., Wang, A.-D., Li, B., Cui, T.-H. & Lu, Y.-F. Electrons dynamics control by shaping femtosecond laser pulses in micro/nanofabrication: modeling, method, measurement and application. *Light: Sci. Appl.* **7**, 17134 (2018).

24. Liu, H., Lin, W. & Hong, M. Hybrid laser precision engineering of transparent hard materials: challenges, solutions and applications. *Light: Science & Applications* **10**, 1–23 (2021).

25. Bhuyan, M. *et al.* Ultrafast laser nanostructuring in bulk silica, a "slow" microexplosion. *Optica* **4**, 951–958 (2017).

26. Chanal, M. *et al.* Crossing the threshold of ultrafast laser writing in bulk silicon. *Nature Communications* **8**, 1–6 (2017).

27. Götte, N. *et al.* Temporal airy pulses for controlled high aspect ratio nanomachining of dielectrics. *Optica* **3**, 389–395 (2016).

28. Hua, J.-G., Liang, S.-Y., Chen, Q.-D., Juodkazis, S. & Sun, H.-B. Free-form micro-optics out of crystals: Femtosecond laser 3D sculpturing. *Advanced Functional Materials* 2200255 (2022).

29. Sun, K. *et al.* Three-dimensional direct lithography of stable perovskite nanocrystals in glass. *Science* **375**, 307–310 (2022).

30. Juodkazis, S. *et al.* Laser-induced microexplosion confined in the bulk of a sapphire crystal: evidence of multimegabar pressures. *Phys. Rev. Lett.* **96**, 166101 (2006).

31. Vailionis, A. *et al.* Evidence of superdense aluminium synthesized by ultrafast microexplosion. *Nature Communications* **2**, 1–6 (2011).

32. Bellouard, Y. *et al.* Stress-state manipulation in fused silica via femtosecond laser irradiation. *Optica* **3**, 1285–1293 (2016).

33. Lin, H., Jia, B. & Gu, M. Dynamic generation of Debye diffraction-limited multifocal arrays for direct laser printing nanofabrication. *Optics Letters* **36**, 406–408 (2011).

34. Flamm, D. *et al.* Structured light for ultrafast laser micro-and nanoprocessing. *Optical Engineering* **60**, 025105 (2021).

35. Salter, P. S. & Booth, M. J. Adaptive optics in laser processing. *Light: Science & Applications* **8**, 1–16 (2019).

36. Forbes, A., de Oliveira, M. & Dennis, M. R. Structured light. *Nature Photonics* **15**, 253–262 (2021).

37. Velpula, P. K. *et al.* Spatio-temporal dynamics in nondiffractive Bessel ultrafast laser nanoscale volume structuring. *Laser & Photonics Reviews* **10**, 230–244 (2016).

38. Meyer, R. *et al.* Single-shot ultrafast laser processing of high-aspect-ratio nanochannels using elliptical Bessel beams. *Optics Letters* **42**, 4307–4310 (2017).

39. Couairon, A. & Mysyrowicz, A. Femtosecond filamentation in transparent media. *Physics reports* **441**, 47–189 (2007).

40. Mahmoud Aghdami, K., Rahnama, A., Ertorer, E. & Herman, P. R. Laser nano-filament explosion for enabling open-grating sensing in optical fibre. *Nature Communications* **12**, 1–10 (2021).

41. Kanehira, S., Si, J., Qiu, J., Fujita, K. & Hirao, K. Periodic nanovoid structures via femtosecond laser irradiation. *Nano Letters* **5**, 1591–1595 (2005).

42. Xie, C., Meyer, R., Froehly, L., Giust, R. & Courvoisier, F. In-situ diagnostic of femtosecond laser probe pulses for high resolution ultrafast imaging. *Light: Science & Applications* **10**, 1–13 (2021).

43. Sugioka, K. & Cheng, Y. Ultrafast lasers—reliable tools for advanced materials processing. *Light: Science & Applications* **3**, e149–e149 (2014).

44. Motoyoshi, M. Through-silicon via (TSV). *Proceedings of the IEEE* **97**, 43–48 (2009).

45. Kawata, S., Sun, H.-B., Tanaka, T. & Takada, K. Finer features for functional microdevices. *Nature* **412**, 697–698 (2001).

46. Rapp, L. *et al.* High aspect ratio micro-explosions in the bulk of sapphire generated by femtosecond Bessel beams. *Scientific Reports* **6**, 1–6 (2016).

47. Bhuyan, M. *et al.* High aspect ratio nanochannel machining using single shot femtosecond Bessel beams. *Applied Physics Letters* **97**, 081102 (2010).

48. Li, Z., Allegre, O. & Li, L. Realising high aspect ratio 10 nm feature size in laser materials processing in air at 800 nm wavelength in the far-field by creating a high purity longitudinal light field at focus. *Light: Science & Applications* **11**, 339 (2022).

49. Lin, Z., Liu, H., Ji, L., Lin, W. & Hong, M. Realization of 10 nm features on semiconductor surfaces via femtosecond laser direct patterning in far field and in ambient air. *Nano Letters* **20**, 4947–4952 (2020).

50. Meyer, R. *et al.* Extremely high-aspect-ratio ultrafast Bessel beam generation and stealth dicing of multi-millimeter thick glass. *Applied Physics Letters* **114**, 201105 (2019).

51. Bellouard, Y., Said, A., Dugan, M. & Bado, P. Fabrication of high-aspect ratio, micro-fluidic channels and tunnels using femtosecond laser pulses and chemical etching. *Optics express* **12**, 2120–2129 (2004).

52. Juodkazis, S. *et al.* Control over the crystalline state of sapphire. *Adv. Mater.* **18**, 1361–1364 (2006).

53. Ródenas, A. *et al.* Three-dimensional femtosecond laser nanolithography of crystals. *Nat. Photon.* **13**, 105 (2019).

54. Meyer, R., Giust, R., Jacquot, M., Dudley, J. M. & Courvoisier, F. Submicron-quality cleaving of glass with elliptical ultrafast Bessel beams. *Applied Physics Letters* **111**, 231108 (2017).

55. Bricchi, E., Klappauf, B. G. & Kazansky, P. G. Form birefringence and negative index change created by femtosecond direct writing in transparent materials. *Optics Letters* **29**, 119–121 (2004).

56. Stoian, R. Volume photoinscription of glasses: three-dimensional micro-and nanostructuring with ultrashort laser pulses. *Applied Physics A* **126**, 1–30 (2020).

57. Mishchik, K. *et al.* Ultrafast laser induced electronic and structural modifications in bulk fused silica. *Journal of Applied Physics* **114**, 133502 (2013).

58. Lancry, M. *et al.* Ultrafast nanoporous silica formation driven by femtosecond laser irradiation. *Laser & Photonics Reviews* **7**, 953–962 (2013).

59. Hentschel, M. *et al.* Dielectric Mie voids: confining light in air. *Light: Science & Applications* **12**, 3 (2023).

60. Xu, Q., Almeida, V. R., Panepucci, R. R. & Lipson, M. Experimental demonstration of guiding and confining light in nanometer-size low-refractive-index material. *Optics letters* **29**, 1626–1628 (2004).

61. García-Lechuga, M. *et al.* Simultaneous time-space resolved reflectivity and interferometric measurements of dielectrics excited with femtosecond laser pulses. *Physical Review B* **95**, 214114 (2017).

62. Gualtieri, J. G., Kosinski, J. A. & Ballato, A. Piezoelectric materials for acoustic wave applications. *IEEE Transactions on Ultrasonics, Ferroelectrics, and Frequency Control* **41**, 53–59 (1994).

63. Wei, D. *et al.* Experimental demonstration of a three-dimensional lithium niobate nonlinear photonic crystal. *Nature Photonics* **12**, 596–600 (2018).

64. Nishiura, S., Tanabe, S., Fujioka, K. & Fujimoto, Y. Properties of transparent Ce: YAG ceramic phosphors for white LED. *Optical Materials* **33**, 688–691 (2011).

65. Zhang, J. *et al.* Ultra-wide bandgap semiconductor $Ga_2O_3$ power diodes. *Nature communications* **13**, 3900 (2022).

66. Liao, M., Shen, B. & Wang, Z. Progress in semiconductor $\beta$-$Ga_2O_3$. *Ultra-Wide Bandgap Semiconductor Materials; Elsevier: Amsterdam, The Netherlands* 263–345 (2019).

67. Goldstein, D. H. *Polarized light* (CRC press, 2017).

68. Eaton, S. W., Fu, A., Wong, A. B., Ning, C.-Z. & Yang, P. Semiconductor nanowire lasers. *Nature Reviews Materials* **1**, 1–11 (2016).

69. Liang, Z., Wu, J., Cui, Y., Sun, H. & Ning, C.-Z. Self-optimized single-nanowire photoluminescence thermometry. *Light: Science & Applications* **12**, 36 (2023).

70. Sander, T. *et al.* Magnetoencephalography with a chip-scale atomic magnetometer. *Biomedical optics express* **3**, 981–990 (2012).

71. Andermann, M. L. *et al.* Chronic cellular imaging of entire cortical columns in awake mice using microprisms. *Neuron* **80**, 900–913 (2013).

72. Delgoffe, A., Nazir, S., Hakobyan, S., Hönninger, C. & Bellouard, Y. All-glass miniature GHz repetition rate femtosecond laser cavity. *Optica* **10**, 1269–1279 (2023).

73. Sugioka, K. *et al.* Femtosecond laser 3D micromachining: a powerful tool for the fabrication of microfluidic, optofluidic, and electrofluidic devices based on glass. *Lab on a Chip* **14**, 3447–3458 (2014).

74. Rahim, K. & Mian, A. A review on laser processing in electronic and MEMS packaging. *Journal of Electronic Packaging* **139** (2017).

75. Li, L., Kong, W. & Chen, F. Femtosecond laser-inscribed optical waveguides in dielectric crystals: a concise review and recent advances. *Advanced Photonics* **4**, 024002 (2022).

76. Buschlinger, R., Nolte, S. & Peschel, U. Self-organized pattern formation in laser-induced multiphoton ionization. *Phys. Rev. B* **89**, 184306 (2014).

77. Déziel, J.-L., Dubé, L. J. & Varin, C. Dynamical rate equation model for femtosecond laser-induced breakdown in dielectrics. *Physical Review B* **104**, 045201 (2021).